\documentclass[reprint,aps,prd,nofootinbib,twocolumn,superscriptaddress,preprintnumbers]{revtex4-2}

\usepackage[T1]{fontenc}
\usepackage[usenames,dvipsnames]{xcolor}
\usepackage{aas_macros,hyperref,mathrsfs,orcidlink}
\hypersetup{colorlinks=true,citecolor=red,linkcolor=red,urlcolor=blue}

\usepackage{ulem}

\usepackage{graphicx,physics,bm,amssymb,amsmath,amsthm,amsfonts,times}

\usepackage{amsmath}
\usepackage{enumitem}
\usepackage{esint}

\begin{document}

\title{Distorting Kerr Images with Parity-Odd Scalar Hair}

\author{Qian Wan}
\affiliation{School of Physics and Optoelectronics, South China University of Technology, Guangzhou 510641, P. R. China}

\author{Yehui Hou}
\email{yehuihou@sjtu.edu.cn}
\affiliation{Tsung-Dao Lee Institute, Shanghai Jiao-Tong University, Shanghai 201210, P. R. China}

\author{Yang Huang}
\email{sps_huangy@ujn.edu.cn}
\affiliation{School of Physics and Technology, University of Jinan, Jinan 250022, P. R. China}

\author{Peng-Cheng Li}
\email{pchli2021@scut.edu.cn}
\affiliation{School of Physics and Optoelectronics, South China University of Technology, Guangzhou 510641, P. R. China}

\author{Minyong Guo}
\affiliation{School of Physics and Astronomy, Beijing Normal University, Beijing 100875, P. R. China}
\affiliation{Key Laboratory of Multiscale Spin Physics (Beijing Normal University), Ministry of Education, Beijing 100875, China}

\author{Bin Chen}
\affiliation{Institute of Fundamental Physics and Quantum Technology, \& School of Physical Science and Technology, \\Ningbo University, Ningbo, Zhejiang 315211, P. R. China}
\affiliation{Center for High Energy Physics, Peking University, No.5 Yiheyuan Rd, Beijing 100871, P. R. China}

\begin{abstract}
We investigate thin-disk imaging of Kerr black holes with synchronized scalar hair, focusing on backreacted parity-odd excited states of a complex scalar field minimally coupled to Einstein gravity. 
The scalar energy density is concentrated in two reflection-symmetric, off-equatorial toroidal regions surrounding the black-hole core.
We study the dependence of the images on hair strength and viewing angle, identifying a weak-hair regime close to Kerr. With increasing hair fraction, the photon ring and shadow region shrink and become more distorted. 
In the strong-hair regime, gravitational lensing produces new features, including multiple disconnected shadow components, crescent-shaped structures, and signatures of chaotic lensing.
For nearly edge-on viewing angles, repeated equatorial crossings generate nested ring-like patterns. These results highlight possible geometric signatures of black holes with excited scalar hair.
\end{abstract}

\maketitle

\section{Introduction}
\label{sec1}

Resolving the near-horizon environments of black holes has long been a central goal of gravitational physics. The Event Horizon Telescope (EHT) has now obtained horizon-scale images of the supermassive black holes in M87 and at the Galactic center \cite{EventHorizonTelescope:2019dse,EventHorizonTelescope:2022wkp}. Future instruments with improved dynamic range and angular resolution will further establish black-hole imaging as a probe of strong-field gravity \cite{Johnson:2023ynn, Ayzenberg:2023hfw, Johnson:2024ttr}. Deviations from the Kerr geometry, including those sourced by self-gravitating matter fields, can modify the null-geodesic structure and thereby imprint characteristic features on electromagnetic observables. 

A particularly important class of non-Kerr solutions is provided by Kerr black holes (KBHs) with synchronized scalar hair \cite{Herdeiro:2014goa,Herdeiro:2015gia,Herdeiro:2014ima,Kleihaus:2015iea,Herdeiro:2015tia,Hod:2015bdw,Herdeiro:2016tmi,Delgado:2019prc}, which bifurcate from scalar clouds at the superradiant threshold \cite{Hod:2012px,PhysRevD.90.024051,Benone:2014ssa} and interpolate between vacuum KBHs and rotating boson stars.
Synchronization suppresses the horizon flux of the hairy mode but does not ensure stability against other perturbations; they may remain susceptible to superradiant or nonlinear instabilities, particularly in the strong-hair regime \cite{Ganchev:2017uuo,East:2018glu,Cunha:2019ikd,Carretero:2025eqh,Nicoules:2025bhh}. Nevertheless, stationary ray tracing remains applicable whenever the background evolves slowly compared with the photon-propagation timescale.

The lensing and imaging properties of parity-even KBHs with synchronized scalar hair have been investigated in a variety of settings \cite{Cunha:2015yba,Vincent:2016sjq,Cunha:2016bjh,Gyulchev:2025smf,Gyulchev:2026kvp,Deliyski:2026fav}. 
Related studies have considered rotating hairy or scalarized black holes in other theoretical settings  \cite{Afrin:2021imp,Gyulchev:2024iel,Chen:2026rhr}. 
Parity-odd configurations, corresponding to excited scalar-field states, exhibit a qualitatively different matter distribution: the scalar amplitude vanishes on the equatorial plane, while the energy density is concentrated in two reflection-symmetric, off-equatorial toroidal regions \cite{Wang:2018xhw,Kunz:2019bhm}. 
Through gravitational backreaction, this double-torus structure modifies the horizon geometry, ergoregion, and null-geodesic flow, leaving characteristic imprints on images produced by external illumination.
Under celestial-sphere illumination, parity-odd solutions can produce multiple shadow components and intricate lensing patterns associated with chaotic photon scattering \cite{Huang:2025xqd}. Their appearance under alternative, more astrophysically motivated emission geometries, however, remains largely unexplored.

In this work, we study the optical appearance of KBHs with parity-odd scalar hair illuminated by a geometrically thin accretion disk. 
Disk emission provides a natural probe of the null-geodesic structure in the strong-field region. 
Beyond the multiple shadows and chaotic lensing patterns revealed by celestial-sphere illumination \cite{Huang:2025xqd}, self-lensing in a hairy-KBH--disk system can produce richer image morphologies, including photon-ring substructure, inner shadows, additional dark regions associated with chaotic photon scattering, and characteristic frequency-shift patterns. 

We numerically construct the parity-odd spacetimes by solving the Einstein--Klein--Gordon equations (EKG) and generate their images using backward ray tracing with discrete radiative transfer.
We examine how the photon-ring structure, inner-shadow morphology, and signatures of chaotic photon scattering depend on the hair fraction and observer inclination. 
By comparing with vacuum KBHs, we characterize the principal imaging features associated with the parity-odd configurations considered here.
In the weak-hair regime, the dominant effect is a reduction in the characteristic sizes of the photon ring and inner shadow. 
In the strong-hair regime, the double-torus scalar distribution substantially reorganizes the null-geodesic flow, producing disconnected dark regions, crescent-like features, and localized signatures of chaotic scattering. 
Within the equatorial thin-disk model adopted here, we additionally find multiple disconnected components of the inner shadow, which may provide a potentially parity-sensitive, though model-dependent, imaging feature.

The remainder of this paper is organized as follows. In Sec.~\ref{sec2}, we review the construction of hairy KBHs and summarize the basic properties of the solutions. In Sec.~\ref{sec3}, we introduce the thin-disk model and analyze the resulting image morphology in the weak- and strong-hair regimes. In Sec.~\ref{sec4}, we summarize our main results and discuss their implications. Throughout, we adopt units with $G=c=1$.

\section{BLACK HOLES WITH PARITY-ODD SCALAR HAIR}
\label{sec2}
In this section, we review the basic properties of scalar-hairy black holes and introduce the parity-odd solutions.
We consider Einstein gravity minimally coupled to a complex scalar field $\Psi(x^\mu)$ with mass $\mu$, described by the action
\begin{equation}
\label{action}
S=\int \mathrm{d}^4x \sqrt{-g}\left(\frac{R}{16\pi}-\nabla^\mu\Psi^*\nabla_\mu\Psi-\mu^2\Psi^*\Psi\right),
\end{equation}
where $R$ is the Ricci scalar. Variation of the action yields the EKG equations
\begin{equation}
\label{eom}
\begin{aligned}
&R_{\mu\nu}-\frac{1}{2}g_{\mu\nu}R=8\pi T_{\mu\nu}, \quad (\nabla^\mu\nabla_\mu-\mu^2)\Psi=0 \,,\\
&T_{\mu\nu}=2\nabla_{(\mu}\Psi^*\nabla_{\nu)}\Psi - g_{\mu\nu}\left(\nabla^\alpha\Psi^*\nabla_\alpha\Psi+\mu^2\Psi^*\Psi\right)\,.
\end{aligned}
\end{equation}
To construct stationary and axisymmetric solutions, we adopt the metric ansatz \cite{Kunz:2019bhm}
\begin{equation}
\label{eq:metricansatz}
\begin{aligned}
\mathrm{d}s^2= &-N_1\,\mathrm{e}^{f_0}\mathrm{d}t^2+N_2\,\mathrm{e}^{f_1}(\mathrm{d}r^2+r^2\mathrm{d}\theta^2) \\
&+N_2\,\mathrm{e}^{f_2}r^2\sin^2\theta\,(\mathrm{d}\varphi-W\,\mathrm{d}t)^2,
\end{aligned}
\end{equation}
where $f_i(r,\theta)$ ($i=0,1,2$) and $W(r,\theta)$ are unknown functions, and
\begin{equation}
N_1(r) =\left(\frac{r-r_h}{r+r_h}\right)^2,\quad 
N_2(r) =\left(1+\frac{r_h}{r}\right)^4.
\end{equation}
Here $r_h \geq 0$ denotes the horizon location.
The scalar field is assumed to take the bound-state form
\begin{equation}
\label{eq:fieldansatz}
\Psi(x^\mu) =\phi(r,\theta)\,\mathrm{e}^{\mathrm{i}(m\varphi-\omega t)},
\end{equation}
with frequency $\omega<\mu$ and azimuthal number $m\in\mathbb{Z}^+$. Backreaction generically makes $\phi(r,\theta)$ non-separable.
Regularity at the horizon and the absence of scalar flux imply the synchronization condition \cite{Herdeiro:2014goa}
\begin{equation}
\label{eq:synch}
\omega = m\, \Omega_h\,, 
\end{equation}
where $\Omega_h$ is the horizon angular velocity. 
This condition ensures that the scalar field is invariant along the horizon generator $\chi\!=\!\partial_t+\Omega_h\partial_\varphi$, i.e.\ $\mathcal{L}_\chi \Psi\!=\!0$, so that no scalar flux crosses the horizon. Physically, it corresponds to the threshold of superradiance \cite{Press:1972zz, Brito:2015oca}: modes with $\omega<m\Omega_h$ are amplified, while those with $\omega>m\Omega_h$ are absorbed, and only the synchronized case allows stationary bound states.
At spatial infinity, asymptotic flatness requires the metric to approach Minkowski spacetime, while the scalar field decays as
\begin{equation}
\phi \sim r^{-\alpha} e^{-\sqrt{\mu^2-\omega^2}\,r},
\end{equation}
reflecting the localized bound state. The power-law factor encodes subleading Coulomb-type corrections and does not affect the leading exponential falloff.

To solve the EKG equations, we impose boundary conditions ensuring regularity of the metric and scalar field at the event horizon and symmetry axis, together with asymptotic flatness at spatial infinity (see \cite{Huang:2024gtu,Huang:2025xqd} for more details). The boundary condition at the equatorial plane is fixed by the parity of the scalar field. For parity-odd configurations,
\begin{equation}
\phi(r,\theta)=-\phi(r,\pi-\theta)
\quad\Rightarrow\quad
\phi(r,\pi/2)=0\,,
\end{equation}
such that the equatorial plane is a nodal surface of the scalar field. Despite the antisymmetry of the amplitude, the energy-momentum tensor is invariant under equatorial reflection because it depends on combinations quadratic in the field and its derivatives. Consequently, the scalar energy density is displaced away from the equatorial plane and concentrated in two reflection-symmetric, off-equatorial toroidal regions \cite{Wang:2018xhw}.

This nontrivial angular structure modifies the multipolar distribution of the matter content and, consequently, the spacetime geometry, in contrast to parity-even configurations where the scalar field typically peaks at the equator. Such differences are expected to affect the effective potential for geodesic motion, providing a distinctive imprint on strong-field observables.

\begin{figure}[htbp]
	\includegraphics[width=3in]{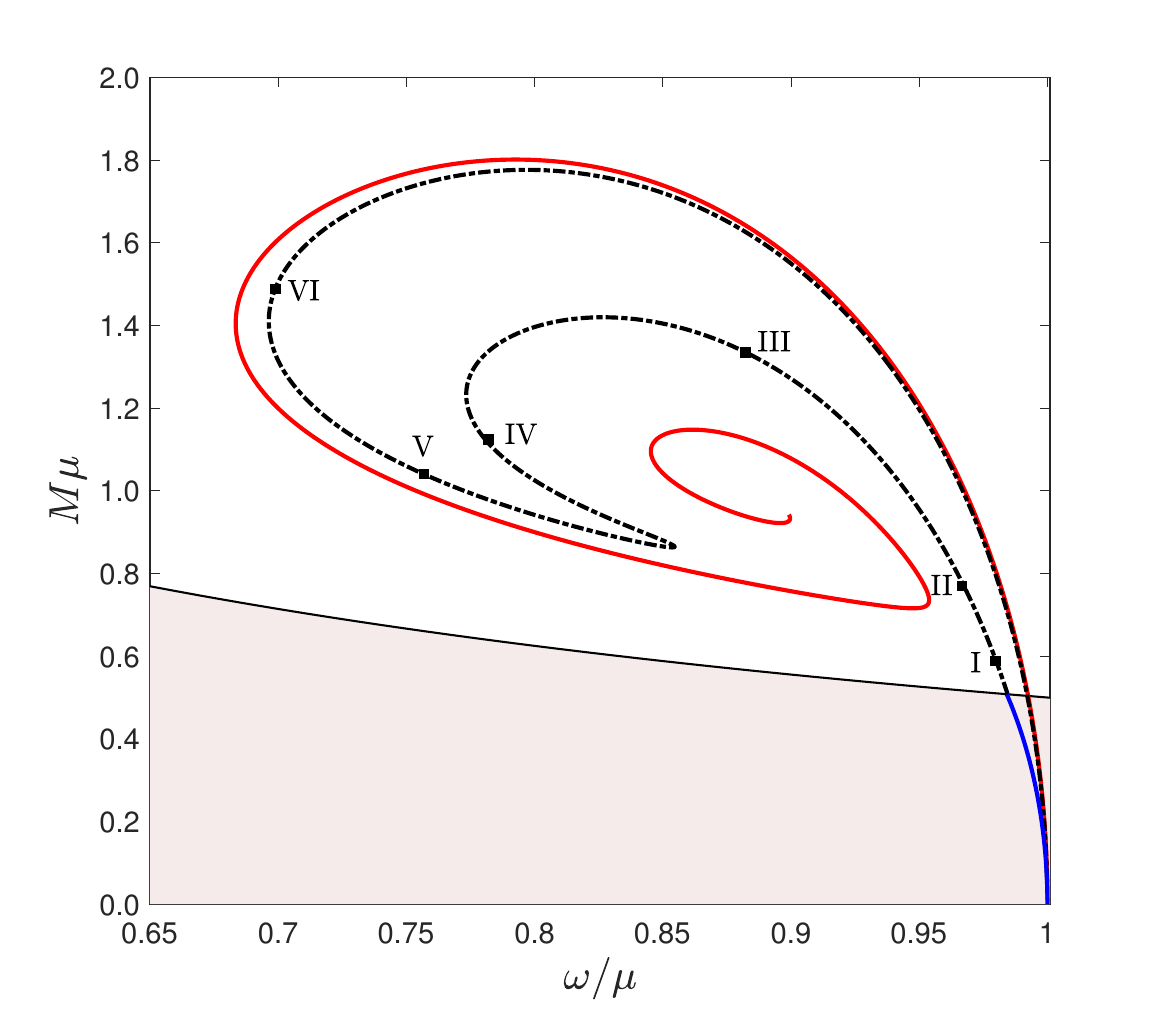}
	\caption{Parameter space of parity-odd KBHs with scalar hair ($m=1$) in the $M-\omega$ plane. The light pink region represents the vacuum Kerr solutions, while the solid black curve marks the extremal KBH boundary. The solid red and blue curves represent the boson-star and scalar-cloud solutions, respectively. The six black squares mark the parity-odd hairy black-hole solutions used below for thin-disk imaging. All lie on the family with $r_h\!=0.01 \mu^{-1}$, shown by the black dash-dotted curve; their parameters are listed in Table~\ref{tab1}.}
	\label{figpara}
\end{figure}

The spacetime is characterized by conserved global charges: the ADM mass $M$ and total angular momentum $J$, together with a Noether charge associated with the global $U(1)$ symmetry,
\begin{equation}
Q_{\Psi} = \int \mathrm{d}^3\bm{x}\sqrt{-g}\,j^0, \quad
j^{\mu}=\mathrm{i}\bigl(\Psi\nabla^\mu\Psi^*-\Psi^*\nabla^\mu\Psi\bigr).
\end{equation}
For stationary, axisymmetric solutions described by Eq.\eqref{eq:metricansatz}, Eq.\eqref{eq:fieldansatz}, the ADM mass and angular momentum admit the decomposition
\begin{equation}
M = M_h + M_{\Psi}, \quad J = J_h + J_{\Psi},
\end{equation}
where $M_{\Psi}$, $J_{\Psi} \!=\! m Q_{\Psi}$ are the scalar-field contributions \cite{Herdeiro:2014goa}. The quantities $M_h$ and $J_h$, defined by Komar integrals on the horizon, satisfy the Smarr relation $M_h = 2T_h S + 2\Omega_h J_h$ \cite{Herdeiro:2015gia}. To quantify the scalar contribution, we further introduce
\begin{equation}
q = J_{\Psi}/J,
\end{equation}
which measures the fraction of angular momentum stored in the scalar field. For co-rotating solutions, $q\in[0,1]$: the Kerr limit is recovered as $q\to 0$, while $q\to 1$ approaches a boson-star configuration \cite{Kaup:1968zz,Ruffini:1969qy}.

\begin{table}
	\begin{center}
\caption{Physical quantities (in units of $\mu$) of the hairy black hole solutions. Each configuration is labeled as $X_q^{r_h}$, where the Roman numeral identifies the configuration number, the subscript $q$ denotes the hair fraction, and the superscript $r_h$ specifies the horizon radius.}
		\begin{tabular}{cccccccc}
			\toprule
			Label & $\omega$ & $M$ & $J$ & $q$ & $M_\Psi/M$ & $J/M^2$ & $J_h/M_h^2$ \\
			\hline
			$\text{\uppercase\expandafter{\romannumeral1}}_{0.304}^{0.01}$ & $0.980$ & $0.588$ & $0.334$ & $0.304$ & $0.175$ & $0.966$ & $0.988$ \\
			$\text{\uppercase\expandafter{\romannumeral2}}_{0.566}^{0.01}$ & $0.967$ & $0.770$ & $0.510$ & $0.566$ & $0.383$ & $0.860$ & $0.981$ \\
			$\text{\uppercase\expandafter{\romannumeral3}}_{0.821}^{0.01}$ & $0.882$ & $1.334$ & $1.122$ & $0.821$ & $0.711$ & $0.630$ & $1.355$ \\
			$\text{\uppercase\expandafter{\romannumeral4}}_{0.923}^{0.01}$ & $0.782$ & $1.125$ & $0.857$ & $0.923$ & $0.886$ & $0.677$ & $4.028$ \\
			$\text{\uppercase\expandafter{\romannumeral5}}_{0.997}^{0.01}$ & $0.757$ & $1.041$ & $0.778$ & $0.997$ & $0.973$ & $0.718$ & $2.977$ \\
			$\text{\uppercase\expandafter{\romannumeral6}}_{0.999}^{0.01}$ & $0.699$ & $1.489$ & $1.409$ & $0.999$ & $0.979$ & $0.636$ & $1.376$ \\
			\toprule
		\end{tabular}
		\label{tab1}
	\end{center}
\end{table}

Fig.~\ref{figpara} presents the parameter space for the $m=1$ branch, where the red and black solid lines represent the boson stars and extremal KBHs, respectively, and the light pink region corresponds to the vacuum Kerr solutions. 

The black dash-dotted curve denotes the hairy KBHs for $\mu r_h\!=\!0.01$, which start from the Minkowski limit at $\omega\!=\!\mu$ and terminate at the scalar cloud solutions represented by the solid blue line.
Decreasing the horizon radius enlarges the domain of existence of black-hole solutions. 
We thus select six representative configurations with fixed $r_h \!=\! 0.01 \mu^{-1}$ (black squares) for subsequent analysis. 
These solutions are denoted by $X_{q}^{r_h}$, where $X$ labels the solution, $q$ is the scalar angular-momentum fraction, and $r_h$ is the horizon radius. The corresponding physical parameters are listed in Table~\ref{tab1}.

There is a paucity of studies on the stability of parity-odd solutions in the literature. Here, we assess their mechanical stability using a Newtonian toy model. In the large-hair-fraction regime, we treat the black hole as a test particle located at the center of the scalar distribution, which is modeled by two identical thin rings of radius $R$, with linear mass density $\chi$, positioned at $z=\pm h$. In cylindrical coordinates $(\rho,z,\varphi)$, the corresponding gravitational potentials are
\begin{equation}
    \Phi_\pm(\rho,z)
    =-\int_0^{2\pi}
    \frac{\chi R\,\dd\tilde{\varphi}}
    {\sqrt{\rho^2+R^2-2R\rho\cos\tilde{\varphi}+(z\mp h)^2}}\,.
\end{equation}
Near the origin, the total potential $\Phi=\Phi_++\Phi_-$ is
\begin{equation}
    \Phi\approx
    -\frac{4\pi R\chi}{\sqrt{R^2+h^2}}
    +\frac{\pi R\chi(2h^2-R^2)}
    {(R^2+h^2)^{5/2}}(\rho^2-2z^2)\,.
\end{equation}
Thus, except when $2h^2\!=\!R^2$, the origin is a saddle point, precluding simultaneous radial and vertical stability.  
This suggests that strongly hairy parity-odd configurations may be vulnerable to off-centering perturbations.

Nonlinear evolutions of parity-even hairy black holes yield lifetimes of $\gtrsim 10^{4}M$ for weakly hairy configurations and $10^{2}M$--$10^{3}M$ near the boson-star limit \cite{Carretero:2025eqh,Nicoules:2025bhh}. These estimates cannot be directly extrapolated to the present solutions based solely on their scalar-mass and angular-momentum fractions.
Nevertheless, even the shortest of these timescales is longer than the strong-field light-crossing time, $\tau_{\rm lens}\sim 1M$--$10M$, thereby supporting the use of stationary ray tracing for most lensing observables. 
For configurations IV--VI, however, the evolution timescale may become comparable to the residence time of highly chaotic photon trajectories, potentially modifying the resulting fine-scale fractal structure.

\section{Disk Imaging}
\label{sec3}

In this section, we study the image of the hairy black hole illuminated by an equatorial disk. To extract main morphological features, we adopt a geometrically thin, optically thin disk with a synthetic emissivity profile. This simplified setup captures the essential lensing signatures while minimizing astrophysical uncertainties.

\begin{figure*}[htbp]
	\centering
	\includegraphics[width=6in, trim=5 5 5 5, clip]{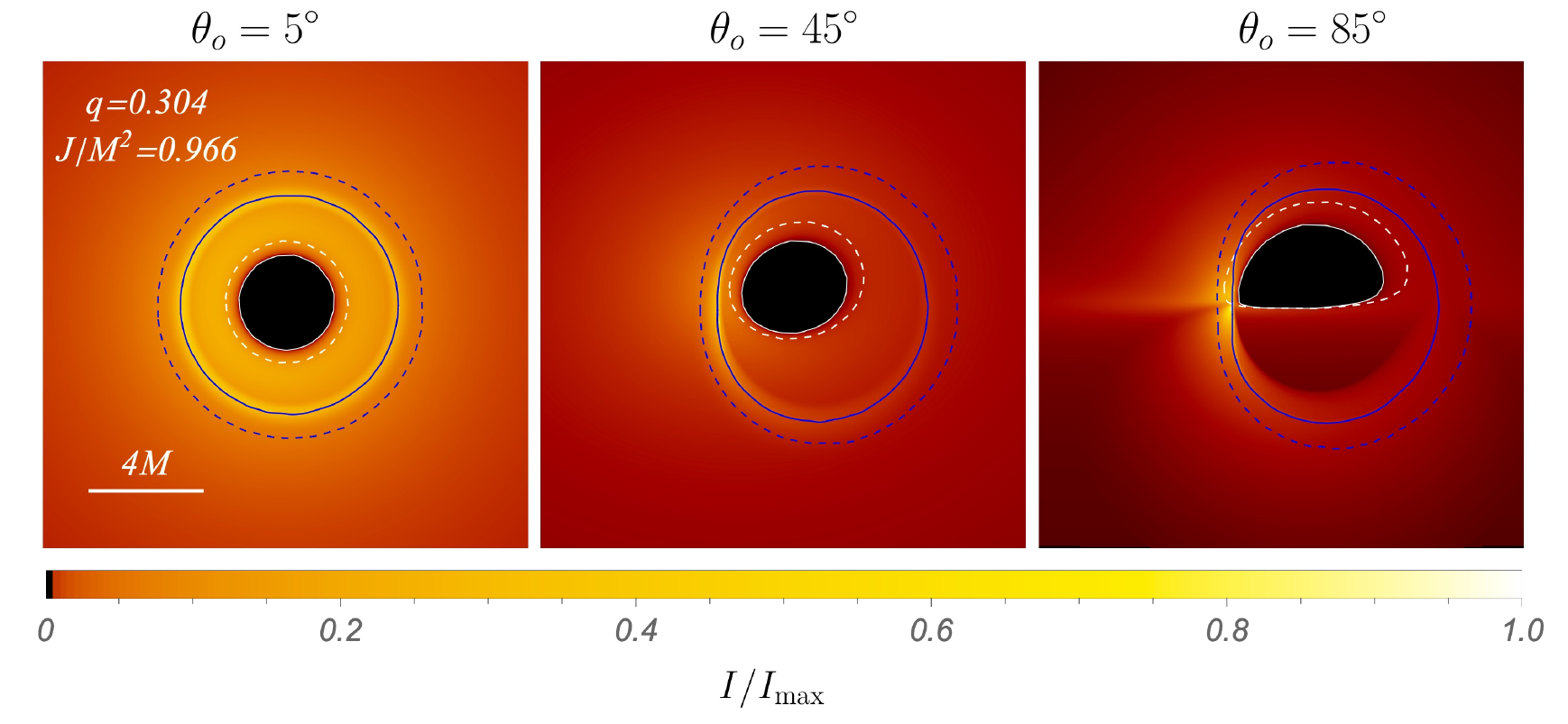}
\caption{Thin-disk images of $\text{\uppercase\expandafter{\romannumeral1}}_{0.304}^{0.01}$ at different viewing angles. The solid white and blue curves denote the photon ring and the inner-shadow boundary, respectively. For comparison, the dashed white and blue curves indicate the corresponding boundaries of the photon ring and inner shadow for a vacuum Kerr black hole with the same ADM mass and angular momentum as configuration $\text{\uppercase\expandafter{\romannumeral1}}_{0.304}^{0.01}$. The observer is fixed at a radial distance of $100M$.}
	\label{figsmallq}
\end{figure*}

\subsection{Emission profile}

We model the accretion flow as a stationary, equatorial disk extending from the horizon to an outer radius $r_{\rm out}=50M$, beyond which the emission is negligible. For the class of solutions considered here, the equatorial circular geodesics admit an innermost stable circular orbit (ISCO) outside the horizon. Outside the ISCO, matter moves on Keplerian geodesics with four-velocity $u^\mu=u^t(1,0,0,\Omega_{\rm K})$, where
\begin{equation}
	\begin{aligned}
		&\Omega_{\rm K}=\frac{-\partial_r g_{t\phi}+\sqrt{(\partial_r g_{t\phi})^2-\partial_r g_{\phi\phi}\,\partial_r g_{tt}}}{\partial_r g_{\phi\phi}}\,, 
	\end{aligned}
\end{equation}
and $u^t=\left(-g_{\phi\phi}\Omega_{\rm K}^2-2g_{t\phi}\Omega_{\rm K}-g_{tt}\right)^{-1/2}$, following from the normalization and circular geodesic equations \cite{chandrasekhar1985mathematical}.
Inside the ISCO, stable circular motion ceases and the plasma plunges toward the black hole. 
We approximate this region by a ballistic flow that conserves the specific energy $E_{\rm ISCO}$ and angular momentum $L_{\rm ISCO}$ inherited from the ISCO \footnote{This kinematic prescription approximates the plunging flow as an equatorial timelike geodesic with conserved $E_{\rm ISCO}$ and $L_{\rm ISCO}$, neglecting pressure gradients, magnetic stresses, and dissipation. It yields a continuous velocity field across the ISCO for computing frequency shifts but does not capture the detailed plasma dynamics.}:
\begin{equation}
	\begin{aligned}
		&u^i=-g^{ti}E_{\rm ISCO}+g^{\phi i}L_{\rm ISCO}\,,\quad i = t,\phi \\
		&u^r=-\sqrt{\frac{g_{tt}(u^t)^2+2g_{t\phi}u^t u^\phi+g_{\phi\phi}(u^\phi)^2+1}{-g_{rr}}}\,,
	\end{aligned}
\end{equation}
where the negative sign of $u^r$ corresponds to inward motion.

Photons emitted from the disk are strongly lensed before reaching a distant observer and may intersect the equatorial plane multiple times. Each intersection contributes independently to the observed intensity. Using the Lorentz-invariant relation $I_\nu/g^3\!=\!\text{const.}$ along null geodesics \cite{Lindquist:1966igj}, the total observed specific intensity can be written as
\begin{equation}
	I_{\nu_o}=\sum_{n=1}^{N_\text{max}} g_n^3 f_n J_n\,,
\end{equation}
where $n$ labels the $n$-th equatorial crossing, $g_n=\nu_o/\nu_n$ is the redshift factor, and $J_n$ is the local emissivity evaluated at the emission point. The factor $f_n$ accounts for the incident angle dependence; for simplicity we set $f_n=1$, which has been shown not to qualitatively affect the image morphology \cite{Gralla:2019xty}.

For the emissivity, we adopt a synthetic, isotropic profile of the form
\begin{equation}
	\ln J(r) = p_1\ln\left(r/M\right)+p_2\left[\ln\left(r/M\right)\right]^2,
\end{equation}
with parameters $p_1=-2$ and $p_2=-1/2$ \cite{Chael:2021rjo}. This phenomenological model reproduces the radially concentrated emission expected in realistic accretion flows.

Although simplified, the disk prescription can capture main features of black hole images \cite{Hou:2022eev, Zhang:2023bzv, Meng:2025ivb, He:2026odk, Ou:2026ohc}. More sophisticated effects, such as finite disk thickness or anisotropic emission \cite{Hou:2023bep, Zhang:2024lsf, Zhou:2025moa, Wan:2025gbm}, primarily modify fine structures without altering the shadow and photon ring. Thus, the qualitative features discussed below are expected to be robust.

\subsection{Image Morphology}

We compute black hole images using a standard backward ray-tracing scheme. To enable a consistent comparison across different solutions, the observer is fixed as a zero-angular-momentum observer located at $r_o = 100M$.

\subsubsection{Small hair fraction regime}

We begin with the small hair fraction regime. Fig.~\ref{figsmallq} shows the images of $\text{\uppercase\expandafter{\romannumeral1}}_{0.304}^{0.01}$ for several viewing angles. The overall morphology closely resembles that of a vacuum KBH: an asymmetric central dark region surrounded by a narrow bright ring. The bright ring corresponds to the photon ring, formed by photons that orbit the black hole and cross the emission region multiple times \cite{Gralla:2019xty}, while the central dark region represents the direct image of the equatorial horizon (the inner edge of the disk), often referred to as the inner shadow \cite{Chael:2021rjo}.

A direct comparison with the corresponding vacuum KBH (same ADM mass $M$ and angular momentum $J$) shows that a moderate hair fraction primarily induces an overall rescaling of the image, without significantly altering its detailed morphology. As the inclination angle $\theta_o$ increases, both the hairy and vacuum cases exhibit the same qualitative evolution: the photon ring transitions from nearly circular to a characteristic ``D-shape'', while the inner shadow becomes progressively flattened into a hat-like structure.

Stronger scalar hair could modify the size and shape of the rings and shadows.
To quantify these differences, we introduce geometric diagnostics for both the photon ring and the inner shadow. For a given closed convex curve, we define its centre on the image plane as
$x_c=(x_\text{max}+x_\text{min})/2$ and $y_c=(y_\text{max}+y_\text{min})/2$, 
where the subscripts ``max'' and ``min'' denote the maximum and minimum values of the abscissa (ordinate). 
By adopting polar coordinates $(\rho,\alpha)$ centred at $(x_c,y_c)$, the relative deviation from a reference vacuum KBH is defined as \cite{Johannsen:2013vgc,Cunha:2015yba}
\begin{equation}
	\sigma_i^\text{Kerr}=\sqrt{\frac{1}{2\pi}\int_0^{2\pi}\left[\frac{\rho_i(\alpha)-\rho_i^\text{Kerr}(\alpha)}{\rho_i^\text{Kerr}(\alpha)}\right]^2\dd\alpha}\,,
\end{equation}
where $i=\mathrm{PR}$ and $\mathrm{IS}$ denote the photon ring and inner shadow, respectively. We also define the average radius as 
\begin{equation}
	\bar{\rho}_i=\frac{1}{2\pi}\int_0^{2\pi}\rho_i(\alpha)\dd\alpha\,.
\end{equation}
The results are summarised in Table~\ref{tab2}, where two reference Kerr spacetimes are considered: one with the same ADM quantities $(M,J)$ (labelled by ``ADM''), and another with the same horizon quantities $M_h$ and $J_h = (1-q)J$ (labelled by ``Hori''). This distinction allows us to disentangle the contributions from the total spacetime versus the near-horizon geometry.

\begin{table*}
	\centering
	\begin{tabular}{ccccccccccccccc}
		\toprule 
		& & $\theta_o$  & $\quad$ & $\bar{\rho}_\text{PR}$ & $\bar{\rho}_\text{PR}^\text{ADM}$ & $\sigma_\text{PR}^\text{ADM}(\%)$ & $\bar{\rho}_\text{PR}^\text{Hori}$ &  $\sigma_\text{PR}^\text{Hori}(\%)$ & $\quad$ & $\bar{\rho}_\text{IS}$ & $\bar{\rho}_\text{IS}^\text{ADM}$ & $\sigma_\text{IS}^\text{ADM}(\%)$ & $\bar{\rho}_\text{IS}^\text{Hori}$ & $\sigma_\text{IS}^\text{Hori}(\%)$ \\
		\hline
		& & $5^\circ$ & & $3.926$ & $4.816$ & $18.48$ & $3.965$ & $1.023$ & & $1.707$ & $2.284$ & $25.26$ & $1.809$ & $5.656$ \\
		$\text{\uppercase\expandafter{\romannumeral1}}_{0.304}^{0.01}$ & & $45^\circ$ && $3.989$ & $4.869$ & $18.12$ & $4.001$ & $0.887$ & & $1.768$ & $2.335$ & $24.28$ & $1.852$ & $4.635$ \\
		&& $85^\circ$ && $4.011$ & $4.938$ & $18.87$ & $4.059$ & $1.645$ & & $1.953$ & $2.544$ & $23.22$ & $2.034$ & $4.069$ \\
		\hline
		&& $5^\circ$ && $2.991$ & $4.901$ & $38.97$ & $2.969$ & $1.145$ & & $1.295$ & $2.502$ & $48.26$ & $1.373$ & $5.809$ \\
		$\text{\uppercase\expandafter{\romannumeral2}}_{0.566}^{0.01}$ & & $45^\circ$ && $3.127$ & $4.949$ & $36.89$ & $3.004$ & $4.459$ & & $1.381$ & $2.542$ & $46.65$ & $1.407$ & $2.279$ \\
		&& $85^\circ$ && $3.056$ & $5.007$ & $39.01$ & $3.046$ & $1.324$ & & $1.485$ & $2.712$ & $45.31$ & $1.539$ & $4.174$ \\
		\hline
		&& $5^\circ$ && $1.817$ & $5.039$ & $63.95$ & -- & -- & & $0.716$ & $2.714$ & $73.62$ & -- & -- \\
		$\text{\uppercase\expandafter{\romannumeral3}}_{0.821}^{0.01}$ && $45^\circ$ && $2.186$ & $5.054$ & $56.83$ & -- & -- & & $1.011$ & $2.743$ & $63.08$ & -- & -- \\
		&& $85^\circ$ && $1.760$ & $5.085$ & $65.42$ & -- & -- & & $0.851$ & $2.874$ & $70.85$ & -- & -- \\
		\toprule
	\end{tabular}
	\caption{Average radii and relative deviations of the photon ring (PR) and the inner shadow (IS) for $\text{\uppercase\expandafter{\romannumeral1}}_{0.304}^{0.01}$, $\text{\uppercase\expandafter{\romannumeral2}}_{0.566}^{0.01}$, and $\text{\uppercase\expandafter{\romannumeral3}}_{0.821}^{0.01}$, together with those of the corresponding vacuum KBHs. The superscripts ``ADM'' or ``Hori'' denote vacuum KBHs matched by ADM or horizon quantities, respectively. The observer is located at a radial distance of $100M$}
	\label{tab2}
\end{table*}

The third to seventh columns present the photon-ring results. The average ring radius decreases monotonically with increasing hair fraction. Its inclination dependence, however, changes qualitatively with $q$. Only $\text{\uppercase\expandafter{\romannumeral1}}_{0.304}^{0.01}$ exhibits the Kerr-like monotonic behavior, whereas for $\text{\uppercase\expandafter{\romannumeral2}}_{0.566}^{0.01}$ and $\text{\uppercase\expandafter{\romannumeral3}}_{0.821}^{0.01}$, $\bar{\rho}_{\rm PR}$ initially increases and then decreases with $\theta_o$. The deviation from a KBH with the same ADM quantities also increases with $q$. As a larger fraction of the total mass and angular momentum is stored in the scalar cloud, the central horizon becomes smaller than that of the corresponding KBH, reducing both the photon ring and the inner shadow.

By contrast, the deviation from a KBH with matching horizon quantities remains small. It is approximately $1\%$ in all cases except $\text{\uppercase\expandafter{\romannumeral2}}_{0.566}^{0.01}$ at $\theta_o=45^\circ$, suggesting that the photon-ring size is primarily governed by the near-horizon geometry. The larger deviation of $4.459\%$ in this case results from the non-monotonic inclination dependence of $\bar{\rho}_{\rm PR}$, which is associated with the scalar-cloud distribution, as discussed below. For $\text{\uppercase\expandafter{\romannumeral3}}_{0.821}^{0.01}$, the Kerr spacetime with the same horizon quantities would satisfy $J_h>M_h^2$ and hence correspond to a naked singularity; it is therefore excluded from the comparison.

The last five columns summarize the inner-shadow properties. Its mean radius generally decreases with increasing $q$, reflecting the shrinking central dark region. 
For $\text{\uppercase\expandafter{\romannumeral3}}_{0.821}^{0.01}$, however, $\bar{\rho}_{\rm IS}$ and $\sigma_{\rm IS}^{\rm ADM}$ vary non-monotonically with $\theta_o$, unlike in Kerr. This behavior is associated with the off-equatorial energy-density peaks of the double-torus scalar distribution, which modify the relevant null geodesics. For $\text{\uppercase\expandafter{\romannumeral2}}_{0.566}^{0.01}$, $\bar{\rho}_{\rm IS}$ increases monotonically with $\theta_o$, whereas $\sigma_{\rm IS}^{\rm Hori}$ first decreases and then increases owing to the different inclination dependences of the hairy and horizon-matched Kerr inner shadows.
Except for $\text{\uppercase\expandafter{\romannumeral2}}_{0.566}^{0.01}$ at $\theta_o=45^\circ$, the inner shadow deviates more strongly than the photon ring and is therefore generally more sensitive to scalar hair.

\subsubsection{Large hair fraction regime}

\begin{figure*}[htbp]
	\centering
	\includegraphics[width=5.8in, trim=60 146 60 190, clip]{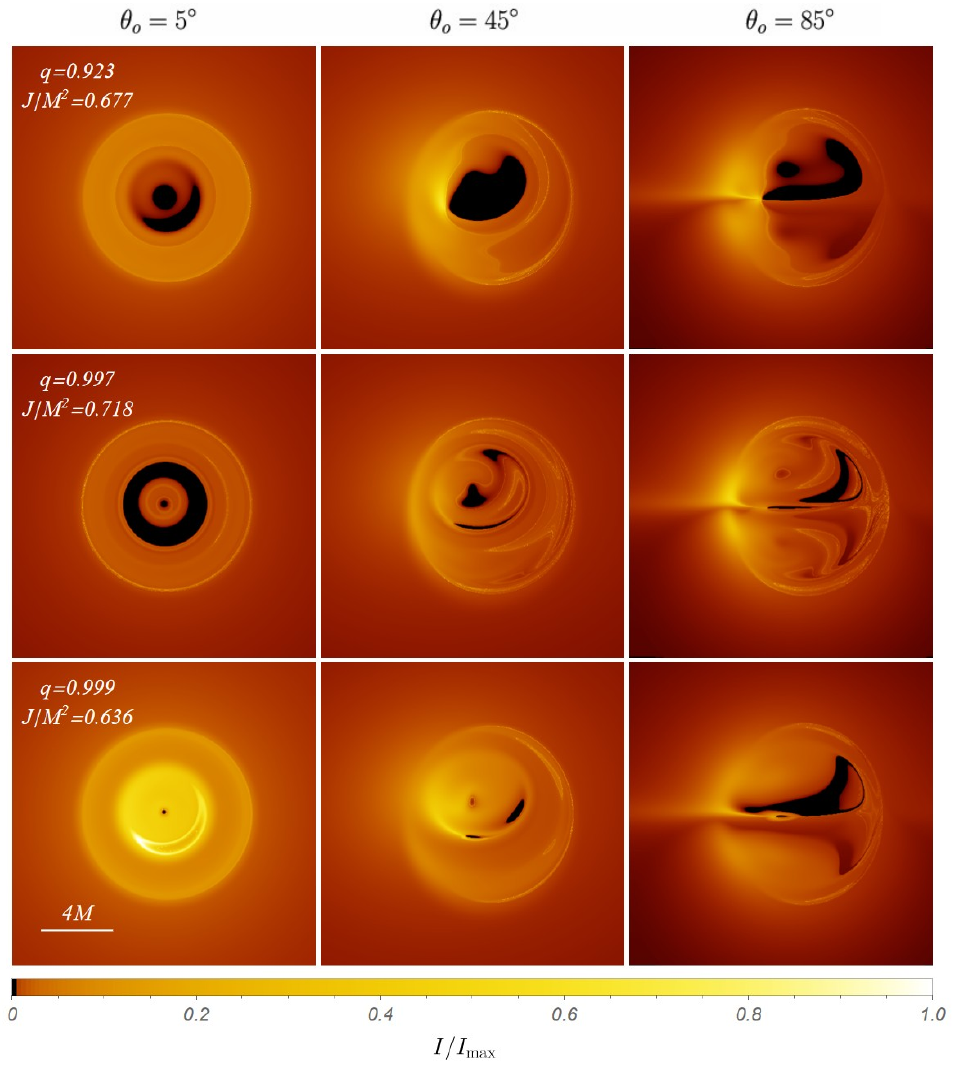}
	\caption{Thin-disk images at different viewing angles. From top to bottom, the hairy black holes are $\text{\uppercase\expandafter{\romannumeral4}}_{0.923}^{0.01}$, $\text{\uppercase\expandafter{\romannumeral5}}_{0.997}^{0.01}$ and $\text{\uppercase\expandafter{\romannumeral6}}_{0.999}^{0.01}$, respectively. The corresponding  parameters can be found in Table \ref{tab1}.}
	\label{figlargeq}
\end{figure*}

\begin{figure*}[htbp]
\hspace{4mm}
\includegraphics[width=6.4in]{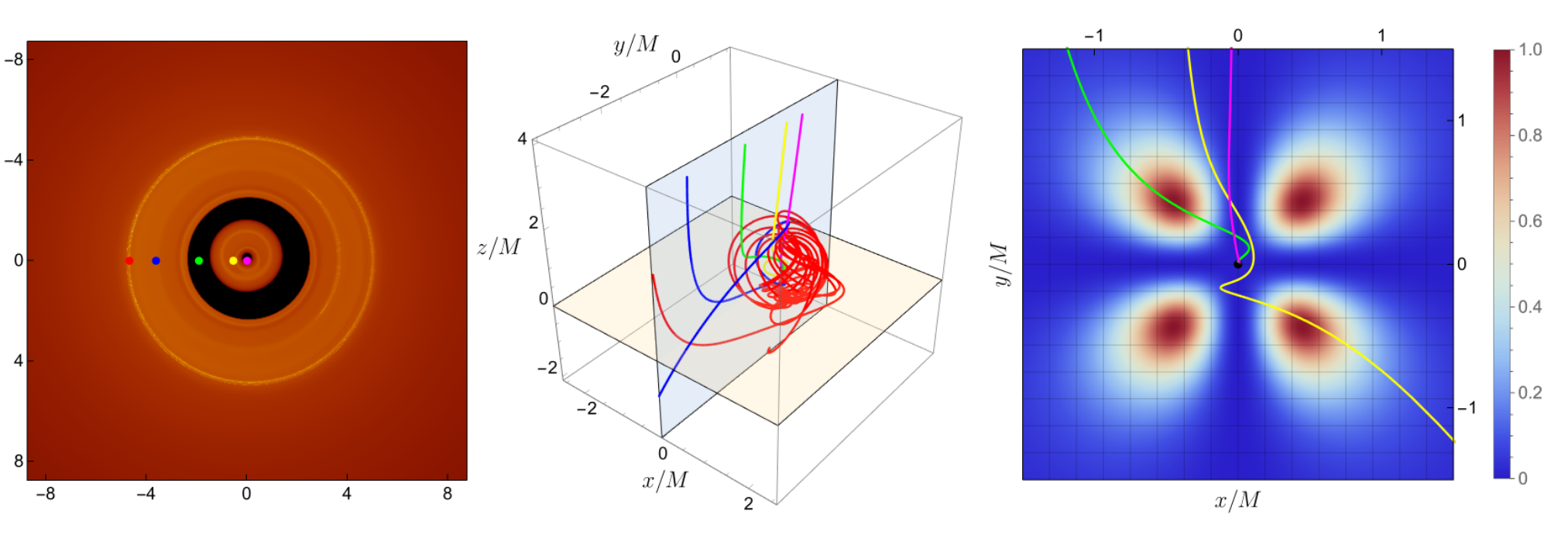}\\
\hspace{1.6mm}
\includegraphics[width=6.6in]{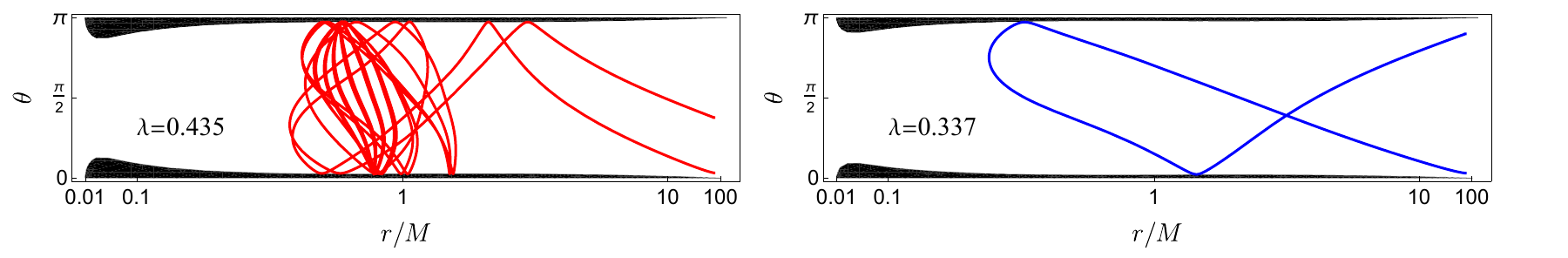}
\caption{Illustration of light trajectories corresponding to five selected points in the black hole image for configuration $\text{\uppercase\expandafter{\romannumeral5}}_{0.997}^{0.01}$ at a viewing angle $\theta_o = 5^\circ$. \textbf{Top left:} locations of these points on the image plane. \textbf{Top middle:} a three-dimensional schematic, where the light-yellow and light-blue planes denote the equatorial plane and the $yz$-plane, respectively. We adopt a Cartesian-like coordinate system $(x,y,z)$ centered on the black hole, constructed from $(r,\theta,\phi)$ as if they were standard spherical coordinates. \textbf{Top right:} density map of the squared scalar-field amplitude $|\phi(r,\theta)|^2$ on the $yz$-plane, together with the projected trajectories of the selected light rays. 
\textbf{Bottom:} trajectories of two representative rays (corresponding to the red and blue points) projected onto the $r$--$\theta$ plane, where $\lambda$ is the impact parameter of the photons, and the black regions indicate the forbidden region defined by the effective potential \cite{Cunha:2016bjh}.}
	\label{figlens}
\end{figure*}

As the hair fraction increases to sufficiently large values, the resulting images exhibit qualitative deviations from those of vacuum Kerr black holes. In Fig.~\ref{figlargeq}, we show images for three representative solutions ($\text{\uppercase\expandafter{\romannumeral4}}_{0.923}^{0.01}$, $\text{\uppercase\expandafter{\romannumeral5}}_{0.997}^{0.01}$, and $\text{\uppercase\expandafter{\romannumeral6}}_{0.999}^{0.01}$), all with $q>0.9$, viewed from different inclination angles. As shown in Table~\ref{tab1}, for these configurations, most of the mass and angular momentum are carried by the scalar field. Owing to its parity-odd structure, the orbital dynamics and photon trajectories are strongly altered, giving rise to disconnected shadow regions and signatures of chaotic scattering, similar to that reported in \cite{Gyulchev:2026kvp}.

For nearly face-on views (left column of Fig.~\ref{figlargeq}), the central shadow shrinks as $q$ increases. At the same time, a secondary dark region emerges at larger radii, evolving from a crescent into a ring, and eventually giving way to a bright crescent-like structure as $q\to 1$ (bottom-left panel). We have verified that both disconnected dark regions correspond to direct projections of the equatorial horizon, i.e., null geodesics that connect the equatorial horizon to the observer without re-crossing the equatorial plane. In the horizonless limit (boson star), one expects the inner shadow to disappear for typical accretion models \cite{Olivares:2018abq, Rosa:2023qcv}. 
Compared to the celestial-sphere illumination results shown in Fig.~\ref{figcs} in the Appendix, disk imaging provides additional information about strong lensing by such hairy black holes.

To elucidate the origin of the crescent-like structure, we analyze photon trajectories associated with selected points on the observer’s screen for the solution $\text{\uppercase\expandafter{\romannumeral5}}_{0.997}^{0.01}$ at $\theta_o=5^\circ$ (Fig.~\ref{figlens}). The trajectories exhibit qualitatively distinct behaviors. Some photons (e.g., red curve) cross the equatorial plane multiple times, indicating the presence of unstable photon orbits \cite{Gralla:2019xty}. Others (purple and green curves) are captured by the horizon, forming the inner and outer shadow regions. In contrast, photons in between (yellow curve) escape to infinity, producing bright structures between the shadows.

Overlaying these trajectories with the scalar field distribution $|\phi(r,\theta)|^2$ reveals that the scalar hair induces strong, spatially inhomogeneous lensing. In particular, it can deflect trajectories that would otherwise miss the observer (e.g., green curve), while frame dragging further distorts photon paths (yellow curve), leading to multiple inflection points in their projections. This interplay between scalar-field-induced lensing and rotation underlies the formation of disconnected shadows and complex image morphology.

\begin{figure*}[htbp]
\hspace{-5mm}
\includegraphics[width=6.4in]{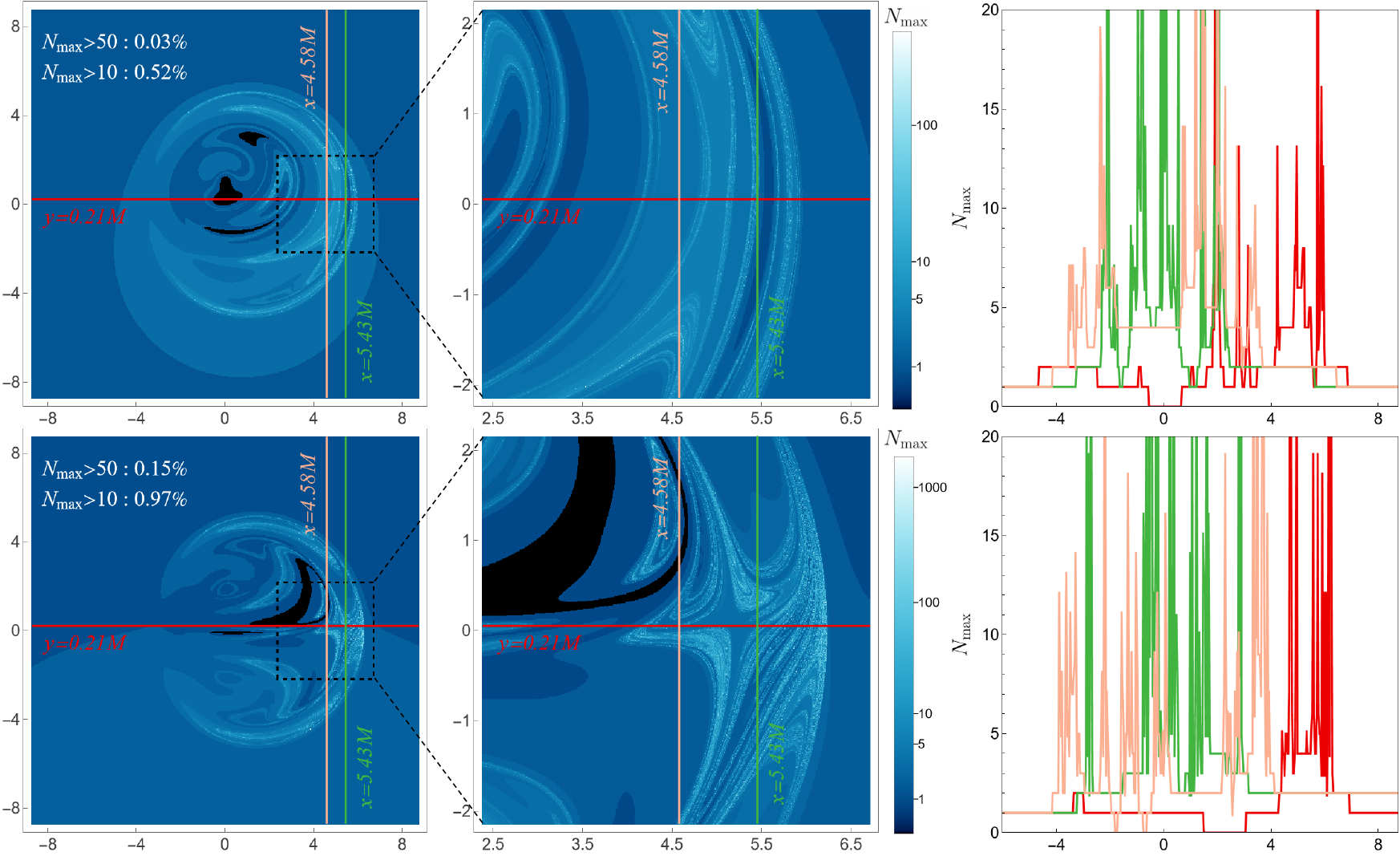}
\caption{Statistical maps of the total number of equatorial crossings, $N_{\rm max}$, for light rays in the images of $\mathrm{\uppercase\expandafter{\romannumeral5}}_{0.997}^{0.01}$. \textbf{Left:} Images at $\theta_o=45^\circ$ and $85^\circ$, with the fractions of photons satisfying $N_{\rm max}>10$ and $N_{\rm max}>50$ indicated in the upper-left corners. \textbf{Middle:} Enlarged views of the regions outlined by the black dashed boxes. \textbf{Right:} One-dimensional profiles of $N_{\rm max}$ along three representative lines, marked by different colors in the left panels. The irregular behavior of $N_{\rm max}$ provides clear signatures of chaos \cite{Shipley:2016omi,Bacchini:2021fig,Wang:2022kvg,Berens:2026qoh}.}
	\label{figchao}
\end{figure*}

As the viewing angle increases, the image becomes progressively more distorted: shadow boundaries develop irregular shapes, and merging or splitting of shadow components can occur. Simultaneously, relativistic Doppler boosting enhances emission from approaching regions of the disk and suppresses receding ones, producing a pronounced brightness asymmetry. In addition, some images exhibit irregular, fine-scale brightness patterns, which are indicative of chaotic scattering, i.e., a sensitive dependence of photon trajectories on their initial conditions \cite{Wang:2022kvg, Berens:2026qoh}.

To quantify this behavior, we construct pixel-resolved maps of the maximal number of equatorial crossings, $N_{\max}=0,1,2,\dots$, for photons reaching the observer, neglecting frequency shifts (Fig.~\ref{figchao}). We also report the fractions of photons with $N_{\max}>10$ and $N_{\max}>50$. The middle panel shows a zoomed-in region, while the right panel presents $N_{\max}$ profiles along selected directions.

These maps reveal the presence of nested, ring-like structures composed of photon trajectories undergoing multiple equatorial crossings. Their boundaries are highly intricate and correlate closely with the shadow and bright ring features seen in the images, suggesting that they originate from nearby families of unstable photon orbits undergoing repeated winding. In some regions, photons cross the equatorial plane hundreds or even thousands of times, indicating strong oscillatory motion.
The fraction of high-crossing trajectories increases with the viewing angle, implying that inclined observers probe regions where photons are more likely to undergo repeated near-equatorial oscillations. Nevertheless, even for $\theta_o=85^\circ$, fewer than $1\%$ of photons satisfy $N_{\max}>10$, indicating that these chaotic features remain localized.

The zoomed-in view further shows that the nested structures persist under magnification, continuously splitting into finer filamentary patterns with interwoven regions of high and low $N_{\max}$. Consistently, the profiles in the right panel exhibit sharp peaks and rapid oscillations within narrow intervals, while remaining small elsewhere. This prominent irregularity and strong sensitivity to initial conditions provide clear evidence for localized photon  chaotic scattering.

We briefly assess how dynamical instabilities may affect the images. For an estimated lifetime $\tau_{\rm life}\gtrsim10^{2}M$, the spacetime evolves much more slowly than direct and indirect rays propagate ($N_{\max}=1,2$), leaving the photon ring and inner shadow largely unaffected. 
Highly lensed rays execute repeated oscillations near a narrow radial region around $r_g=M$ (see the red curve in Fig.~\ref{figlens}), which have a characteristic propagation time $\tau_{\rm lens}\simeq N_{\max}\pi r_g$. Thus, most fine structures in configurations IV--VI, generated by rays with $N_{\max}\lesssim10$, should persist. Only features associated with substantially larger $N_{\max}$ may be smeared out, but their image contribution is negligible.

\section{Summary}
\label{sec4}

We have investigated the thin-disk images of KBH with parity-odd scalar hair, 
examining how the image morphology changes as the hair fraction increases. For moderate hair fraction $q$, the hairy solutions largely preserve the lensing structure of their Kerr counterparts, with the dominant effect being a reduction in the sizes of the photon ring and inner shadow. 
As $q$ becomes sufficiently large, the image morphology undergoes a qualitative transition. In this regime, the scalar cloud dominates the exterior geometry and acts as an additional lens outside the horizon. The resulting core--double-halo lensing structure reorganizes the null geodesic flow and gives rise to disconnected shadow regions, chaotic ring-like patterns, and, near the extremal hairy limit, the near disappearance of the central shadow. 

Relative to illumination by a celestial sphere, disk emission partially obscures the shadow and can reduce the apparent size of the dark region. 
However, the combined effects of gravitational lensing and gravitational and Doppler redshifts produce a richer image morphology, including pronounced brightness asymmetries, bright arcs, and elongated dark bands. 
For small $q$, the image morphologies of both parity-odd and parity-even configurations \cite{Gyulchev:2025smf} remain close to that of a KBH. 
At sufficiently large $q$, both parity sectors exhibit non-Kerr lensing signatures, including distorted shadow boundaries, disconnected shadow components, and localized filamentary structures associated with chaotic photon scattering \cite{Cunha:2016bjh,Gyulchev:2025smf}. In the parity-odd solutions studied here, these filamentary features arise from the double-torus scalar distribution, and exhibit a pronounced dependence on the observer inclination.

Within the equatorial thin-disk model adopted here, a potentially parity-sensitive feature is the occurrence of multiple disconnected inner-shadow components. As illustrated in Fig.~\ref{figlens}, these components are direct images of the equatorial horizon, formed by null geodesics that reach the observer without re-crossing the emitting plane. Their formation may be influenced by the off-equatorial scalar distribution. Nevertheless, because the inner-shadow topology also depends on the emission model, we do not claim that this feature is unique to parity-odd solutions. A controlled comparison between the two parity sectors using identical disk and ray-tracing prescriptions is required to assess the robustness of this feature.

Taken together, these results identify a set of imaging signatures that could distinguish parity-odd scalarized black holes from Kerr black holes, especially in the strong-hair regime where the lensing structure is qualitatively modified. A natural next step is to move beyond the idealized thin-disk model considered here and examine these spacetimes in dynamical accretion flows, where time variability and plasma effects may reshape the observable signatures and determine which features remain robust. This direction is reinforced by recent studies showing that time-dependent simulations provide a powerful framework for discriminating among a broad class of non-Kerr spacetimes \cite{Huang:2018rfn,Jiang:2024vgn,Uniyal:2025uvc,Xia:2026lns}. In such magnetized plasma environments, it would also be worthwhile to investigate complementary observables, particularly light curves and spatially resolved polarimetric signatures.

\begin{acknowledgments}
We thank Xinyu Wang and Kai Lin for their assistance. The work is partly supported by NSFC Grant No.12275004, 12205013, 12575048, 12547123, 12588101, 12205104. Y. Huang is also supported by Shandong Provincial Natural Science Foundation (Grant No. ZR2024QA032) and Youth Innovation Group Plan of Shandong Province (Grant No. 2023KJ107). P. Li is also supported by the startup funding of South China University of Technology. M. Guo is also supported by the BNU Tang Scholar.
\end{acknowledgments}

\bibliographystyle{apsrev4-2}
\bibliography{ref}


\appendix

\onecolumngrid

\section{Supplementary Plots}
\label{appdenxi}

\begin{figure*}[htbp]
	\centering
	\includegraphics[width=5.8in, trim=20 140 20 120, clip]{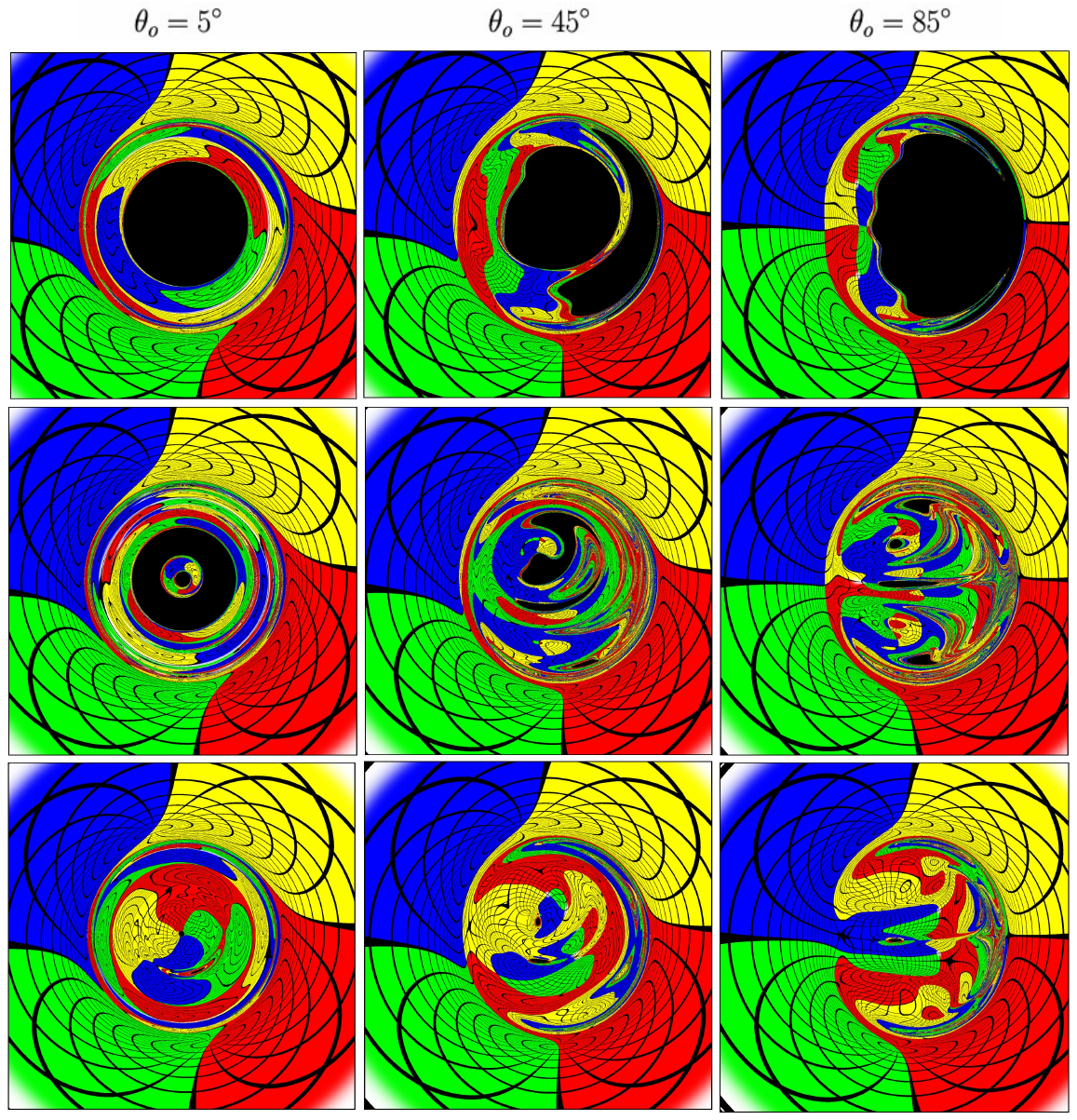}
	\caption{Synthetic images of Kerr black holes with parity-odd scalar hair illuminated by a colored celestial sphere. From top to bottom, the panels correspond to $\text{\uppercase\expandafter{\romannumeral4}}_{0.923}^{0.01}$, $\text{\uppercase\expandafter{\romannumeral5}}_{0.997}^{0.01}$, and $\text{\uppercase\expandafter{\romannumeral6}}_{0.999}^{0.01}$, respectively.}
	\label{figcs}
\end{figure*}

In Fig.~\ref{figcs}, we show images of the black holes \(\text{\uppercase\expandafter{\romannumeral4}}_{0.923}^{0.01}\), \(\text{\uppercase\expandafter{\romannumeral5}}_{0.997}^{0.01}\), and \(\text{\uppercase\expandafter{\romannumeral6}}_{0.999}^{0.01}\) illuminated by a colored celestial sphere introduced in \cite{Cunha:2015yba}. A comparison with Fig.~\ref{figlargeq} indicates that, under thin-disk illumination, the shadow is partially obscured by disk emission and therefore appears significantly smaller. Some features visible in the celestial-sphere setup, such as the crescent-shaped shadow in \(\text{\uppercase\expandafter{\romannumeral5}}_{0.997}^{0.01}\) and the eyebrow-shaped shadow in \(\text{\uppercase\expandafter{\romannumeral6}}_{0.999}^{0.01}\), even disappear entirely. Compared with the celestial-sphere model, the thin-disk setup therefore provides a more stringent criterion for shadow detectability. At the same time, the combined effects of gravitational lensing, disk emission, and gravitational redshift produce a richer image morphology, including strong brightness inhomogeneity, arc-like bright structures, and elongated dark bands. Thin-disk images are thus not only more sensitive to parity-odd scalar hair, but also more relevant to realistic observational situations.

\end{document}